\documentclass[preprint2]{aastex61}
\newcommand\akari{{\itshape AKARI}}
\newcommand\spitzer{{\itshape Spitzer}}
\newcommand\wise{{\itshape WISE}}
\newcommand\src{J051926.26$-$454554.1}
\newcommand\chandra{{\itshape Chandra}}

\received{---}
\revised{---}
\accepted{---}
\shorttitle{Mid-infrared excess from the west hot spot of Pictor A}
\shortauthors{Isobe et al.}

\begin{document} 
%-- Title --%
\title{Mid-infrared excess from the west hot spot of the radio galaxy Pictor A unveiled by \wise}
% \correspondingauthor{Naoki Isobe}
% \email{n-isobe@hp.phys.titech.ac.jp}

%-- authors --%
\author{Naoki Isobe}
\affiliation{
	Institute of Space and Astronautical Science (ISAS), 
        Japan Aerospace Exploration Agency (JAXA), 
        3-1-1 Yoshinodai, Chuo-ku, Sagamihara, Kanagawa 252-5210, Japan}
\email{n-isobe@ir.isas.jaxa.jp}
\affiliation{
	School of Science, Tokyo Institute of Technology, 
	2-12-1 Ookayama, Meguro, Tokyo 152-8551, Japan}
\author{Shoko Koyama}
\affiliation{Academia Sinica Institute of Astronomy and Astrophysics, ,
	PO Box 23-141, Taipei 10617, Taiwan}
\affiliation{
        Max-Planck-Institut f\"{u}r Radioastronomie, 
        Auf dem H\"{u}gel 69, 53121 Bonn, Germany}
\author{Motoki Kino}
\affiliation{
	Kogakuin University, Academic Support Center,
	2665-1 Nakano, Hachioji, Tokyo 192-0015, Japan}
\affiliation{
	National Astronomical Observatory of Japan
       2-21-1 Osawa, Mitaka, Tokyo, 181-8588, Japan}
% \email{motoki.kino@nao.ac.jp}
%\affiliation{
%	Korea Astronomy and Space Science Institute,
%	776 Daedeokdae-ro, Yuseong-gu, Daejeon 305-348, 
%	Republic of Korea} 
\author{Takehiko Wada}
\affiliation{
	Institute of Space and Astronautical Science (ISAS), 
        Japan Aerospace Exploration Agency (JAXA), 
        3-1-1 Yoshinodai, Chuo-ku, Sagamihara, Kanagawa 252-5210, Japan}
\author{Takao Nakagawa}
\affiliation{
	Institute of Space and Astronautical Science (ISAS), 
        Japan Aerospace Exploration Agency (JAXA), 
        3-1-1 Yoshinodai, Chuo-ku, Sagamihara, Kanagawa 252-5210, Japan}
\author{Hideo Matsuhara}
\affiliation{
	Institute of Space and Astronautical Science (ISAS), 
        Japan Aerospace Exploration Agency (JAXA), 
        3-1-1 Yoshinodai, Chuo-ku, Sagamihara, Kanagawa 252-5210, Japan}
\author{Kotaro Niinuma}
\affiliation{
	Graduate School of Science and Engineering, Yamaguchi University, Yoshida
	1677-1, Yamaguchi, Yamaguchi 753-8512, Japan}
\author{Makoto Tashiro}
\affiliation{
	Institute of Space and Astronautical Science (ISAS), 
        Japan Aerospace Exploration Agency (JAXA), 
        3-1-1 Yoshinodai, Chuo-ku, Sagamihara, Kanagawa 252-5210, Japan}
\affiliation{
	Department of Physics, Saitama University, 
        255 Shimo-Okubo, Sakura-ku, Saitama, 338-8570, Japan}

%-- Abstract --%
\begin{abstract}
Mid-infrared properties are reported 
of the west hot spot of the radio galaxy Pictor A
with the Wide-field Infrared Survey Explorer (\wise). 
The mid-infrared counterpart to the hot spot, \wise\ \src,  
is listed in the {\itshape AllWISE} source catalog.
The source was detected in all the four \wise\ photometric bands. 
A comparison between the \wise\ and radio images reinforces 
the physical association of the \wise\ source to the hot spot.
The \wise\ flux density of the source was carefully evaluated.
A close investigation of the multi-wavelength synchrotron spectral energy distribution 
from the object reveals a mid-infrared excess 
at the wavelength of $\lambda=22$ $\mu$m 
with a statistical significance of $4.8 \sigma$
over the simple power-law extrapolation 
from the synchrotron radio spectrum.
The excess is reinforced by single and double cutoff 
power-law modeling of the radio-to-optical spectral energy distribution. 
The synchrotron cutoff frequency of the main and excess components 
was evaluated as $7.1 \times 10^{14}$ Hz and $5.5 \times 10^{13}$ Hz, 
respectively.
From the cutoff frequency,
the magnetic field of the emission region was constrained 
as a function of the region size. 
In order to interpret the excess component,
an electron population different from 
the main one dominating the observed radio spectrum
is necessary.
The excess emission is proposed to originate in a sub structure
within the hot spot, in which the magnetic field is
by a factor of a few stronger than that in the minimum energy condition.
The relation of the mid-infrared excess to the X-ray emission 
is briefly discussed. 
\end{abstract}


%---------------------------------%
%-- Section 1: Introduction --%
%---------------------------------%
\section{Introduction} %==============================
\label{sec:intro}      
Radio galaxies with an FR II morphology \citep{fanaroff74}
usually host compact bright synchrotron radio structures, 
called hot spots \citep[e.g.,][]{begelman84}, 
at the terminal region of their jets. 
It is widely believed that the hot spots correspond to a strong shock 
created by the jet interaction with the ambient intergalactic medium. 
In the hot spots, particles are accelerated 
into the relativistic energies via the diffusive shock acceleration, 
usually called the first-order Fermi process
\citep[e.g.,][]{meisenheimer89,meisenheimer97}.
Thus, they are considered as a possible production site 
of high energy cosmic rays \citep[e.g.,][]{hillas84}.
However, the detailed physical conditions in the hot spots 
are under debate, and then, 
observational constraints are indispensable
to explore the particle acceleration mechanism in detail.

Previously, a combination of radio and X-ray spectral information
has been one of the most valuable probes for the physical properties 
of the hot spots \citep[e.g.,][]{hardcastle04}.
Within a framework of the one-zone particle acceleration, 
%(i.e., a single electron population), 
the hot spots are divided into two classes, 
based on the possible radiation mechanism for the X-ray photons. 
The X-ray spectrum from the high-luminosity hot spots 
is successfully interpreted by the synchrotron-self-Compton (SSC) process  
under a magnetic field close to the equipartition value.
In contrast, the high-frequency tail of the synchrotron emission
tends to significantly contribute to the X-ray spectrum 
of the low-luminosity hot spots. 
However, in order to derive a more definite conclusion,
spectral information in between the radio and X-ray bands 
are usually very useful. 

Mid-infrared observations have a potential to shed light on this issue,
by filling this wide frequency gap. 
A pioneering mid-infrared observation 
of the  FR II radio galaxy Cygnus A with the \spitzer\ observatory 
strengthened the SSC scenario 
for its archetypal high-luminosity hot spots \citep{stawarz07}. 
However, the \spitzer\ upper limit on the mid-infrared flux 
of several low-luminosity hot spots seems inconsistent with
the simple one-zone synchrotron interpretation 
for their radio-to-X-ray spectral energy distribution \citep{werner12}.
Utilizing the mid-infrared data, \citet{kraft07} claimed that 
two independent synchrotron-emitting electron populations 
are required to reproduce the wide-band spectral energy distribution 
from the host spots of the radio galaxy 3C 33. 
The double synchrotron interpretation was invoked 
for the infrared knots of the jet in the quasar 3C 273 \citep{uchiyama06}. 
The multi-component interpretation was applied to 
the polarimetric properties observed with the Atacama Large Millimeter Array 
(ALMA) from kpc-scale structures 
associated with the hot spot of 3C 445  \citep{orienti17}.
Thus, the mid-infrared observations have gradually revealed 
that the acceleration phenomena in the hot spots are 
actually complicated and heterogeneous 
rather than the simple one-zone Fermi acceleration. 

In order to extend the mid-infrared knowledge on the host spots,
the science products from the Wide-field Infrared Survey Explorer 
\citep[\wise;][]{wright10} are expected to be one of the ideal tools. 
\wise\ covers a wide mid-infrared range 
with the four photometric bands at the wavelengths of 
$\lambda = 3.4$ (W1), $4.6$ (W2), $12$ (W3), and $22$ $\mu$m (W4).
Especially, the {\itshape AllWISE} source catalog\footnote{Electrically available at 
{\tt http://irsa.ipac.caltech.edu/cgi-bin/Gator/nph-scan?submit=Select\&projshort=WISE}},
which was constructed by a combination of the data obtained 
in the cryogenic and post-cryogenic \citep{mainzer11} phases, 
is useful,
since it tabulates more than 747 million objects,
thanks to its all-sky coverage with a relatively high sensitivity
(e.g., $54$ $\mu$Jy at $3.4$ $\mu$m).

The FR II radio galaxy Pictor A, 
located at the red shift of $z = 0.035$ \citep{eracleous04}, 
is known to host a distinctive radio hot spot at the edge of its west lobe 
\citep[e.g.,][]{perley97}. 
It is known as a representative of the low-luminosity hot spots 
\citep{hardcastle04}.
Thanks to its high radio intensity, 
the object has been extensively studied not only in the radio band,
but also in the near infrared \citep{meisenheimer97}, optical \citep{thomson95}
and X-ray \citep[e.g.,][]{wilson01,hardcastle16} frequencies.
The high resolution radio image derived with the Very Long Baseline Array (VLBA) 
resolved the hot spot into pc-scale sub structures \citep{tingay08}. 
The hot spot is included in the \spitzer\ hot-spot sample 
by \citet{werner12}. 
However, their result at $24$ $\mu$m 
seems inconsistent to that of \citet{tingay08},
in spite of using the same data set. 
% The result supports the double synchrotron model 
% for its radio-to-X-ray spectral energy distribution,
% where the X-ray emission is radiated from the the sub structures.

In the present paper,
the mid-infrared properties of the west hot spot of Pictor A 
are investigated with \wise.
The mid-infrared spectrum exhibits 
a significant excess over the simple extrapolation of 
its power-law (PL) radio spectrum.  %estimated from \citet{meisenheimer97}.
The excess strongly requires an additional population 
of synchrotron electrons.

Referring to the previous study \citep{tingay08}, 
the cosmological constants are assumed to be 
$H_{\rm 0} = 71$ km s$^{-1}$ Mpc$^{-1}$, 
$\Omega_{\rm m} = 0.27$, and  $\Omega_{\Lambda} = 0.73$
\citep{spergel03}. %, throughout the present paper.
These yield an angle-to-size conversion factor of $688~{\rm pc} / 1\arcsec$
at the rest frame of Pictor A. 

%--------------------% x
% table 1: WISE data % y
%--------------------% z
\begin{deluxetable*}{lccccccc}[t]
\tablecaption{\wise\ properties of the west hot spot of Pictor A.
\label{tab:catalog}}
\tablecolumns{7}
%\tablenum{2}
\tablewidth{0pt}
%==================================================================
\tablehead{ 
	\colhead{Band} & 
	\colhead{$\lambda$ ($\mu$m) \tablenotemark{a}} & 
	\colhead{SN \tablenotemark{b}} & 
	\colhead{$m$ (mag)\tablenotemark{c}} &
	\colhead{$F_{\nu}$ (mJy) \tablenotemark{d}} &
	\colhead{$\sigma_{\rm sys}$ (mJy) \tablenotemark{e}} &
       \colhead{$f_{\rm c}$ \tablenotemark{f}} &
	\colhead{$f_{\rm r}$ \tablenotemark{g}}
	}
%------------------------------------------------------------------
\startdata
	W1 &  3.35 & 45.8 & $13.368 \pm 0.024$ & $1.39 \pm 0.03$ & 0.03 & 0.992 & 1 \\
	W2 &  4.60 & 50.2 & $12.324 \pm 0.022$ & $2.02 \pm 0.04$ & 0.06 & 0.994 & 1 \\
	W3 & 11.56 & 35.7 & $9.569 \pm 0.03$ & $4.60 \pm 0.13$ & 0.21 & 0.937 & 1 \\
	W4 & 22.09 & 13.1 & $7.215 \pm 0.083$ & $9.98 \pm 0.76$ & 0.57 & 0.993 & 0.92 \\
\enddata
%------------------------------------------------------------------
\tablenotetext{a}{The isophotal wavelength of the \wise\ photometric band.}
\tablenotetext{b}{The signal-to-noise ratio.}
\tablenotetext{c}{The source magnitude in the Vega unit.}
\tablenotetext{d}{The corresponding flux density.}
\tablenotetext{e}{The systematic error of the \wise\ photometry \citep{jarrett11}}
\tablenotetext{f}{The color correction factor for $\alpha = 1$.}
\tablenotetext{g}{The additional correction factor for red sources \citep[see][]{wright10}.}
% \tablecomments{}
\end{deluxetable*}

%----------------------------------%
%-- Section 2: Data analysis --%
%----------------------------------%
% \clearpage
\section{Data Analysis}   %==============================
\label{sec:analysis}
\subsection{\wise\ Counterpart to the West Hot Spot of Pictor A} %-----------
\label{sec:WISEcounterpart}
From the {\itshape AllWISE} source catalog,
a moderately bright source, \wise\ \src, was identified 
as the mid-infrared counterpart to the west hot spot of Pictor A.
The \wise\ properties of the source are listed in Table \ref{tab:catalog}.
Among the five photometric measurements in the catalog,
the deep detection profile-fit photometry is adopted. 
The source is detected at all the four \wise\ photometric bands 
with a high signal-to-noise ratio ($>5$). 
The flag information in the catalog indicates that 
the source is point like (${\tt ext\_flg=0000}$),
its photometric quality is high (${\tt ph\_flg=AAAA}$),
it probably exhibited no intensity variation during the \wise\ survey 
(${\tt var\_flg=2111}$),
and it is not contaminated by known artifacts (${\tt cc\_flags=0000}$).

%------------% 
% figure 1 %
%------------%
\begin{figure}[t]
\plotone{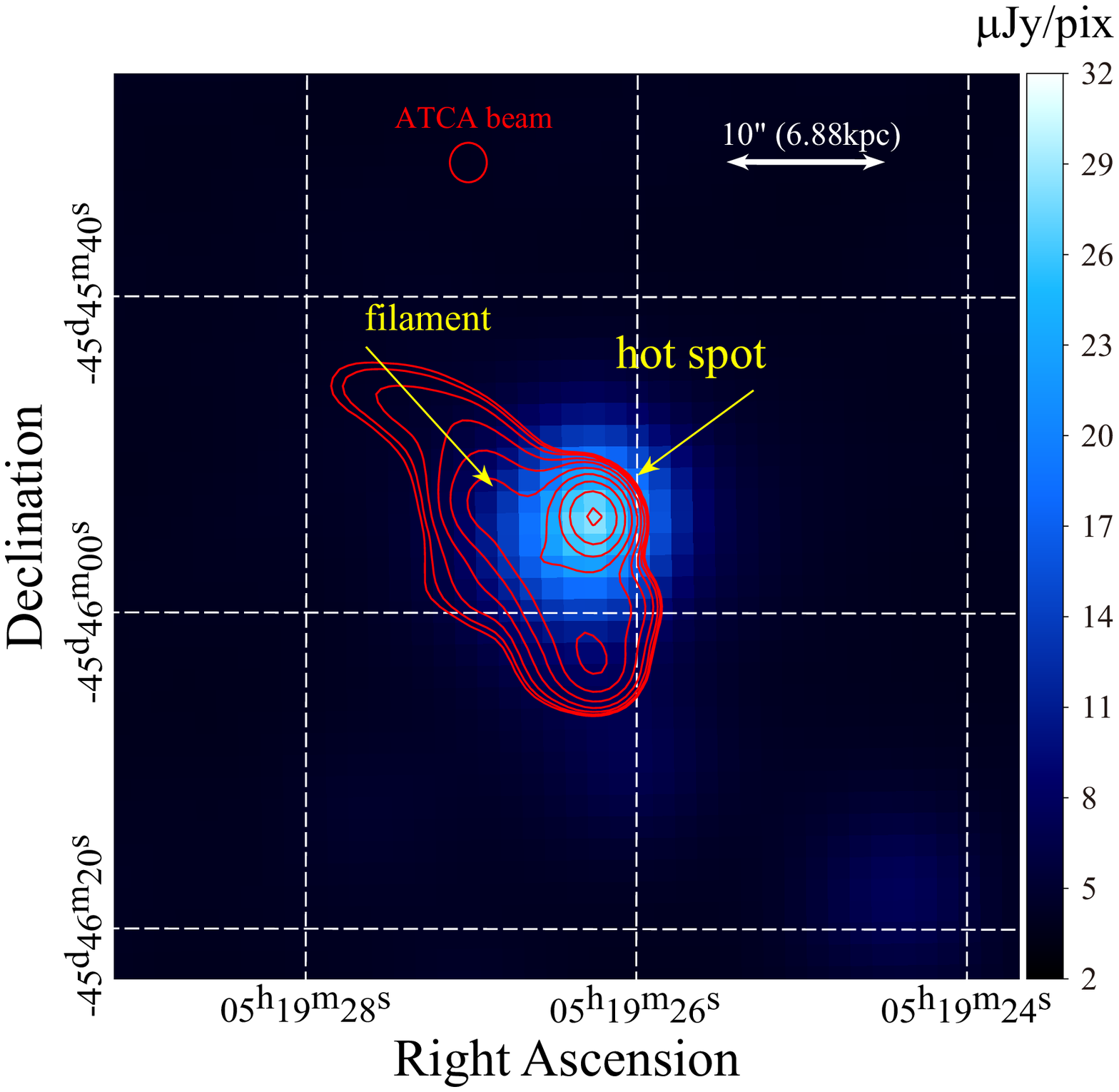}
%{../fig/image/W1_ATCA-4.8GHz_WestHS_cont_beam.eps}
%\vspace{8em}
\caption{3.4 $\mu$m \wise\ image around the west hot spot of Pictor A 
(color scale in the unit of $\mu$Jy pix$^{-1}$).
The pixel size of the image is $1.375\arcsec \times 1.375\arcsec$ 
The 4.8 GHz ATCA image is superposed with contours, 
drawn at 10 levels from 0.018 to 1.8 Jy beam$^{-1}$ in the logarithmic scale.
The ellipse in the top shows the ATCA beam size. 
The hot spot and filament \citep{roser87} are indicated by the arrows.
The horizontal arrow in the top right indicates the angular scale of 10 $\arcsec$,
which corresponds to the physical scale of $6.88$ kpc at the rest frame of Pictor A.
}
\label{fig:image}
\end{figure}

Figure \ref{fig:image} shows the \wise\ image of \wise\ \src\  
at $\lambda = 3.4$ $\mu$m.
The unpublished 4.8 GHz radio image 
obtained with the Australia Telescope Compact Array (ATCA)
(kindly provided by Dr. Lenc; private communication) 
is overlaid with contours. 
The figure supports the physical association of the \wise\ source 
to the hot spot,
rather than to the radio filament on the southeast \citep{roser87}. 
Considering the source density in this field, 
the probability of the chance coincidence is low.

The \wise\ and ATCA coordinates were more precisely registered 
by analyzing the nucleus of Pictor A. 
The \wise\ position of the nucleus is found to agree 
with its ATCA peak within $\sim 20$ mas.
Since this is smaller than the systematic uncertainty of the \wise\ astrometry,
$\sim 50$ mas 
\footnote{See ``Explanatory Supplement to the AllWISE Data Release Products", 
which is electrically available 
at {\tt http://wise2.ipac.caltech.edu/docs/release/allwise/expsup/index.html}},
no astrometric correction was performed.  
Then, the offset of the \wise\ position of the hot spot to its ATCA one 
was evaluated as at most $\sim 300$ mas toward the southeast, 
i.e., the direction to the jet. 
The X-ray position of the hot spot was reported to be shifted 
by $\sim 1\arcsec$ from the radio peak \citep{hardcastle16}, in a similar direction.

\newpage
\subsection{Mid-infrared Spectrum}   %------------------------
\label{sec:WISEspec}
By referring to \citet{wright10}, 
the flux density of the west hot spot 
at the frequency $\nu = \frac{c}{\lambda}$ of the \wise\ photometric band,
is evaluated as 
\begin{equation}
F_{\nu} = \frac{f_{\rm r}}{f_{\rm c}} F_0^* \times 10^{-\frac{m}{2.5}},
\end{equation}
where $c$ is the light speed, $m$ is the \wise\ magnitude,
$F_0^*$ is the zero-magnitude flux density 
for a PL-like spectrum with the energy index of $\alpha = 2$ 
($F_\nu \propto \nu^{-\alpha}$),
$f_{\rm c}$ is the color-correction factor,
and $f_{\rm r}$ is the additional color-correction factor 
for a red source ($\alpha \ge 1$) only applied to the $22$ $\mu$m flux density.
Based on the observed \wise\ color (e.g., $m(3.4)-m(4.6)= 1.04$),
the color correction for $\alpha = 1$ was applied.
The Galactic extinction correction was negligible, 
because it is estimated to be less than $1$\% even at $3.4$ $\mu$m  
from the B-band extinction of $A_{\rm B} = 0.186$ \citep{schlegel85}
and the standard extinction law \citep{rieke85}.
The flux density of the source is summarized in Table \ref{tab:catalog}, 
together with the adopted $f_{\rm c}$ and $f_{\rm r}$ values.
The systematic uncertainty in the \wise\ flux density, $\sigma_{\rm sys}$, 
was based on \citet{jarrett11}.

\begin{figure}[t]
\plotone{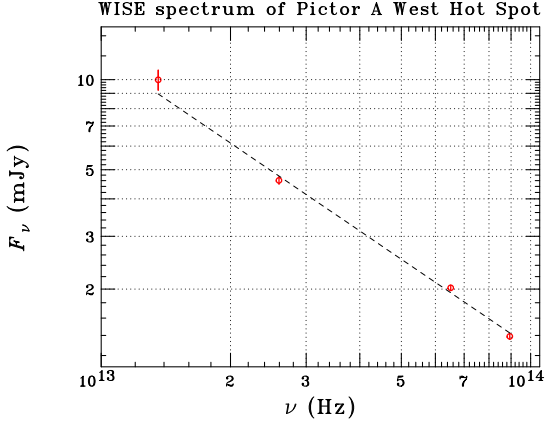}
%{../fig/WISE_spec/WISE_spec.ps}
\caption{\wise\ spectrum of the west hot spot (the open circles).
The best-fit PL model with $\alpha = 0.97$ is plotted with the dashed line.}
\label{fig:WISEspec}。 
\end{figure}

Figure \ref{fig:WISEspec} displays the \wise\ spectrum of 
the west hot spot in $\nu = (1.3$--$9.0)\times 10^{13}$ Hz
(i.e., $\lambda = 3.4$--$22$ $\mu$m).
The spectrum exhibits no break nor cutoff in the \wise\ range. 
It is successfully approximated by a PL model 
(the dashed line in Figure \ref{fig:WISEspec}),
and its index, $\alpha = 0.97 \pm 0.05$, 
becomes self-consistent to the observed \wise\ color,
adopted for the color correction.

%------------% 
% figure 2 %
%------------%

\subsection{Spectral Energy Distribution}  %==============================
\begin{figure}[htbp!]
\includegraphics[width=7.5cm]{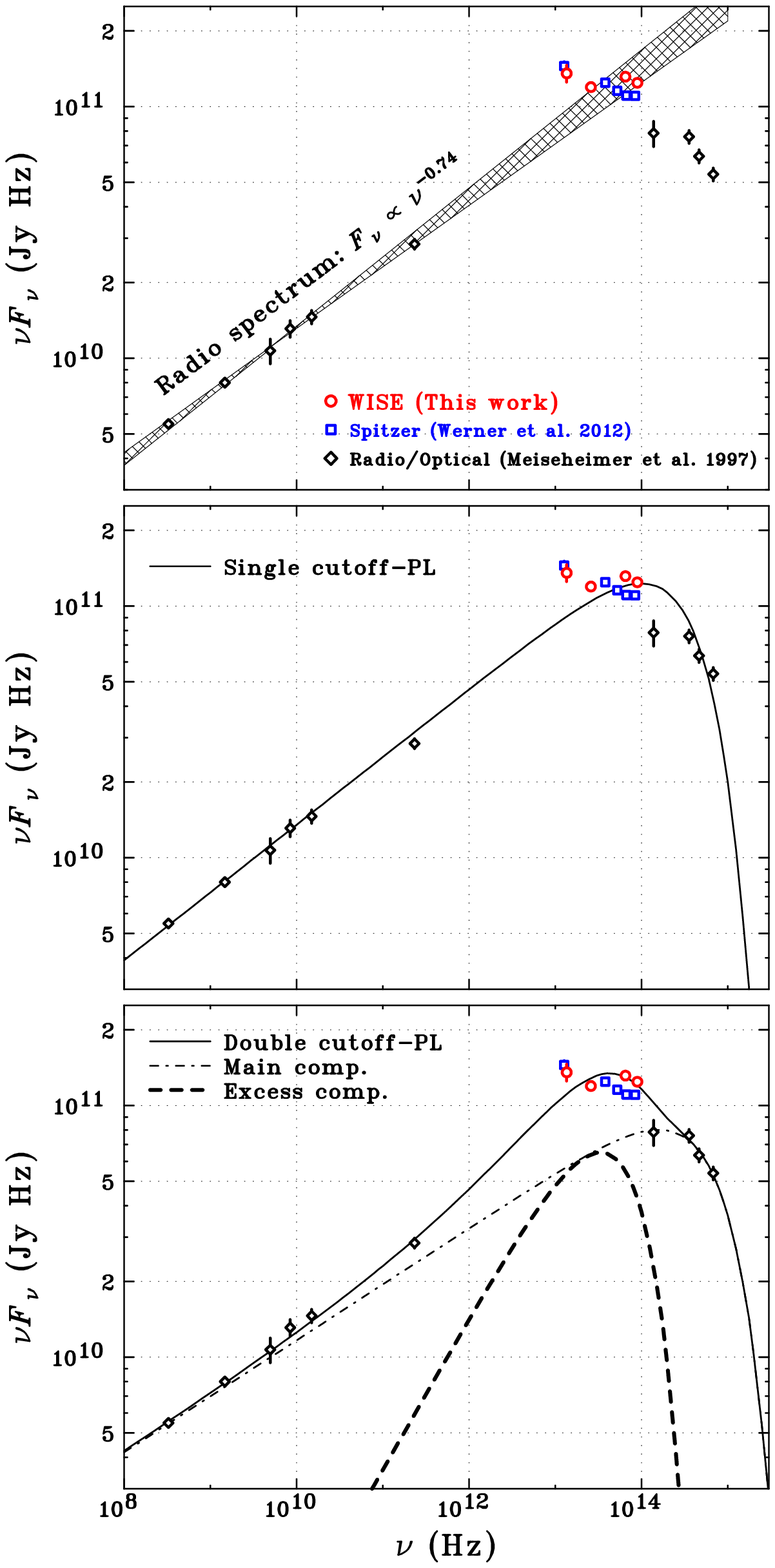}
%\plotone{fig3.ps}
%{../fig/SED/SED_fit.ps}
\caption{Radio-to-optical spectral energy distribution 
of the west hot spot of Pictor A.
The \wise\ data are shown with the circles, 
while the \spitzer\ data \citep{werner12} are plotted with the boxes.
The optical and radio data (the diamonds) are taken from \citet{meisenheimer97}.
On the top panel, the PL model approximating 
the radio data in the 330 MHz--230 GHz range
\citep[$\alpha=0.74\pm0.015$;][]{meisenheimer97} 
is displayed with the hatched tie.
The single cutoff PL model is plotted with the solid line 
in the middle panel. 
The best-fit double cutoff PL model is shown with 
the solid line in the bottom panel.  
The main and excess components are drawn 
with the thin dash-dotted and thick dashed lines, respectively.
}
\label{fig:SED}
\end{figure}

\label{sec:SED}
The spectral energy distribution of the hot spot 
from the radio to optical frequencies 
is compiled in Figure \ref{fig:SED}.
The \wise\ data are displayed with the circles. 
The published data measured with \spitzer\ are also plotted 
with the boxes. 
\citet{tingay08} and \citet{werner12} 
independently reported the results from the same \spitzer\ observation.
Their photometric measurements coincide with each other 
in the $3$--$8$ $\mu$m range within $\sim 10$\%, 
while their results differ at $24$ $\mu$m by $\sim 30$\%,
possibly due to the difference in their photometric aperture.
In the present paper, 
the \spitzer\ spectrum by \citet{werner12} is utilized,
because their $24$ $\mu$m flux is in good agreement 
to the $22$ $\mu$m \wise\ flux, as is seen from Figure \ref{fig:SED}.
The radio, near-infrared, and optical data,
indicated by the diamonds, are take from \citet{meisenheimer97}. 

The object exhibits a straight PL radio spectrum 
over nearly 3 orders of magnitude in the $330$ MHz--$230$ GHz range.
The best-fit flux density at 5 GHz and energy index were 
respectively evaluated as 
$F_\nu = 2.20 \pm 0.02 $ Jy and $\alpha = 0.74 \pm 0.015$
\citep{meisenheimer97}.
As shown with the hatched tie in the top panel of Figure \ref{fig:SED},
its simple extrapolation to the higher frequencies intersects 
the mid-infrared spectrum in the \wise/\spitzer\ frequency range 
around $\nu = 5 \times10^{13}$ Hz (i.e., $\lambda \sim 6$ $\mu$m).
The near-infrared to optical spectrum in the range of $\nu \gtrsim 10^{14}$ Hz
is steeper than the mid-infrared one, indicating a spectral cutoff. 

It is important to note that 
the object shows a PL mid-infrared spectrum with $\alpha \sim 1$
even at the lower frequency range ($\nu < 5 \times10^{13}$ Hz).
Thus, the $22$ $\mu$m \wise\ flux density exceeds 
the extrapolation of the radio spectrum by $3.7$ mJy. 
The significance of this excess is evaluated as $4.8\sigma$.  
Even if the systematic uncertainty of the \wise\ photometry 
(i.e., $\sigma_{\rm sys} = 0.57 $ mJy at $22$ $\mu$m) is taken into account,
the excess remains meaningful ($\sim 3.8\sigma$). 
The result is fully supported by the $24$ $\mu$m \spitzer\ data.  

%------------% 
% figure 3 %
%------------%

%----------------------%x
% table 2: SED fitting %y
%----------------------%z
\begin{deluxetable*}{llll}[t]
\tablecaption{Summary of the spectral modeling to the radio-to-optical data.
\label{tab:SED_fit}}
\tablecolumns{4}
%\tablenum{2}
\tablewidth{0pt}
%==================================================================
\tablehead{ 
	\colhead{Component} & 
	\colhead{Parameter} & 
	\colhead{Single cutoff PL model} & 
	\colhead{Double cutoff PL model} 
	}
\startdata %------------------------------------------------------------------
Main	& $F_{\nu}$(5 GHz) (Jy)  \tablenotemark{a}
				& $2.24 \pm 0.07$ 	& $2.00 \pm 0.14$\\
	& $\alpha$  \tablenotemark{b}		
				& $0.73 \pm 0.01$	& $0.78 \pm 0.02$\\
	& $\nu_{\rm c}$ (Hz)	\tablenotemark{c}
				& $(3.7 \pm 0.4) \times 10^{14}$		 								& $(7.1 \pm 2.2) \times 10^{14}$\\
\hline %------------------------------------------------------
Excess	& $F_{\nu}$(22 $\mu$m) (mJy) \tablenotemark{d}		
				& ---			& $5.1 \pm 1.1$\\
	& $\alpha$  \tablenotemark{b}
				& ---			& $0.40 \pm 0.14$\\
	& $\nu_{\rm c}$ (Hz)	\tablenotemark{c}
				& ---			& $(5.5 \pm 1.8) \times 10^{13}$\\
\hline %------------------------------------------------------
	& $\chi^2/{\rm dof}$	& $6.6$		& $2.3$				\\
\enddata %------------------------------------------------------------------
\tablenotetext{a}{The radio flux density at 5 GHz.}
\tablenotetext{b}{The energy index.}
\tablenotetext{c}{The cut-off frequency.}
\tablenotetext{d}{The mid-infrared flux density at $22$ $\mu$m .}
\end{deluxetable*}

\subsection{Spectral modeling of the mid-infrared excess} %==============================
\label{sec:spec_model}
Firstly, the radio-to-optical spectral energy distribution 
observed from the west hot spot of Pictor A was 
fitted with a single cutoff-PL model. 
For the mid-infrared data, only those with \wise\ were utilized.  
To compensate possible interband flux calibration uncertainties, 
the systematic error on the \wise\ photometry (i.e., $\sigma_{\rm sys}$)
was taken into account. 
The resultant spectral parameters are summarized in Table \ref{tab:SED_fit}. 
The energy index and flux density, 
$\alpha = 0.73 \pm 0.01$ and $F_{\nu} = 2.24 \pm 0.07$ Jy at 5 GHz, 
both agree with those of the radio spectrum in 
the 330 MHz -- 230 GHz range (see \S\ref{sec:SED}).
As shown in the middle panel of Figure \ref{fig:SED},
the fit was rather poor with a reduced chi-square of $\chi^2/{\rm dof}=6.6$.
In addition, the residual of several data points from the model
is found to be larger than $3\sigma$.
It is important to note that 
the mid-infrared excess is basically supported by the single cutoff-PL model.
Its significance at $22$ $\mu$m was evaluated as $3.5 \sigma$ 
when the systematic uncertainty of the \wise\ photometry was considered.

Secondly, 
a double cutoff-PL model was fitted to the observed spectrum,
in order to quantify the excess spectrum.
The main component aims to describe 
the normal synchrotron radiation dominating the radio-to-optical spectrum, 
while the other component is introduced to approximate the mid-infrared excess. 
The best-fit model is shown with the solid line 
in the bottom panel of Figure \ref {fig:SED},
and its spectral parameters are tabulated in Table \ref{tab:SED_fit}.
The observed spectrum was reasonably reproduced 
by the model ($\chi^2/{\rm dof}=2.3$),
and thus, the deviation of the model becomes insignificant 
%($\lesssim 2.7 \sigma$) 
for all the data points.
Therefore, the the double cutoff-PL fitting 
seem mode preferable, rather than the single cutoff-PL one. 

The energy index and cutoff frequency of the main component 
were constrained as $\alpha = 0.78 \pm 0.02$ and 
$\nu_{\rm c} = (7.1 \pm 2.2) \times 10^{14} $ Hz respectively.
The main component becomes slightly softer than the single cutoff-PL model, 
and hence the PL model to the radio spectrum,
probably due to the excess component.  
In the mid-infrared band, 
the excess component (the thick dashed line in Figure \ref {fig:SED}) 
has a similar contribution to that of the main component 
(the dash-dotted line).
The excess component exhibits a hard index of $\alpha = 0.40 \pm 0.14$,
which is consistent to the prediction 
from the strong shock limit ($\alpha = 0.5$).
The cutoff frequency of the excess component,
$\nu_{\rm c} = (5.5 \pm 1.8) \times 10^{13} $ Hz,
is  lower by an order of magnitude than that of the main component.

% \clearpage 
\section{Discussion}      %==============================
\label{sec:discussion}
\subsection{Summary of the mid-infrared excess}
\label{sec:excess}
The west hot spot of the FR II radio galaxy Pictor A 
was identified as the mid-infrared source, \wise\ \src,
listed in the {\itshape AllWISE} source catalog. 
A comparison of the \wise\ and ATCA images confirms that 
the mid-infrared source corresponds to the radio peak of the hot spot
rather than the filamentary structure on the southeast \citep{roser87}.
After carefully evaluating the \wise\ flux density 
(see Table \ref{tab:catalog}),
the mid-infrared energy index of the hot spot 
was evaluated as $\alpha = 0.97 \pm 0.05$.
Figure \ref{fig:SED} indicates that the $22$ $\mu$m \wise\ flux density 
significantly exceeds the simple extrapolation of the radio PL spectrum 
with an index of $\alpha = 0.74 \pm 0.015$.
The excess is reinforced by the single/double cutoff-PL modeling 
of the radio-to-optical spectral energy distribution. 
The mid-infrared spectrum with \spitzer\ \citep{werner12} 
exhibits a similar trend. 

The radio-to-optical spectral energy distribution 
from the west hot spot of Pictor A has been 
widely attributed to synchrotron emission from relativistic electrons
accelerated via the first-order Fermi mechanism in the jet terminal shock
\citep[e.g.,][]{meisenheimer89,meisenheimer97,wilson01,tingay08,werner12}. 
The linear polarization 
in the radio \citep{perley97} and optical \citep{thomson95} frequencies
reinforces the synchrotron origin.
However, the standard one-zone model, 
basically predicting a PL electron energy distribution
with a high-energy cutoff and/or break \citep[e.g.,][]{carilli91},
is unable to reproduce the mid-infrared excess. % uncovered with \wise.
Therefore, 
in order to interpret the mid-infrared data,
a two-zone model is strongly favored.  

Previously, 
mid-infrared properties has been investigated with \spitzer\
for about two dozens of hot spots of FR-II radio galaxies. 
\citet{stawarz07} successfully reproduced 
the radio-to-X-ray spectral energy distribution of the two high-luminosity hot spots
of Cygnus A, namely A and D \citep{hargrave74}, 
with the one-zone synchrotron+SSC model,
where the mid-infrared data agree 
with the high-frequency end of the synchrotron component. 
The mid-infrared spectrum of the radio structures associated 
with the hot spots of the FR II source, 3C 33, 
is reported to be smoothly connected to the synchrotron radio spectrum,
while the second synchrotron component is requested to interpret 
the X-ray spectrum for the most radio structures \citep{kraft07}. 
\citet{werner12} systematically studied the multi-frequency spectrum 
of 17 radio hot spots, 
from which the mid-infrared and X-ray emissions were detected, 
including the west hot spot of Pictor A. 
None of these objects are reported to show a clear sign of the mid-infrared excess. 
However, they utilized the radio data only at 5 GHz instead of 
the wide-band synchrotron radio spectrum, 
and their spectral model is simply based 
on the strong shock assumption ($\alpha = 0.5$).
These simplification are possible to conceal the mid-infrared excess,
since the model with $\alpha = 0.5$ should artificially overpredicts 
the mid-infrared flux density, when it is fitted to the 5 GHz radio data,
especially for sources with a soft radio spectrum ($\alpha > 0.5$). 

Actually, owing to the careful modeling of the synchrotron spectrum,
the present study succeeded in detecting 
the mid-infrared excess from the west hot spot of Pictor A, 
which exhibits a relatively soft radio spectrum ($\alpha=0.74\pm0.015$). 
This result is regarded as the first discovery of the mid-infrared excess 
from hot spots in FR II radio galaxies.

\subsection{How to constrain the magnetic field} %==============================
\label{sec:constraint}
In the following, the two-zone scenario is adopted to examine  
the physical parameters in the hot spot. 
The magnetic field of the main and excess components 
is estimated from the cutoff frequency. 
The cutoff frequency is usually attributed to 
the maximum Lorentz factor of accelerated electrons 
or the break Lorentz factor due to radiative cooling. 
The previous studies on the west hot spot 
of Pictor A \citep[e.g.,][]{tingay08} suggests that 
the synchrotron cooling dominates over the inverse-Compton one. 
The analytical estimates by \citet{kino04} indicates that 
the observed spectral parameters of the object 
are found to be within the cooling break regime.

By equating the synchrotron cooling time scale 
$t_{\rm syn} = \frac{ 3 m_{\rm e} c}{ 4 u_{\rm B} \sigma_{\rm T} \gamma_{\rm b} }$
to the adiabatic loss time scale $t_{\rm ad} = \frac{R} {v} $,
the break Lorentz factor is derived as 
$\gamma_{\rm b} = \frac{6 \pi m_{\rm e} v c }{\sigma_{\rm T} B^{2} R}$
\citep[e.g.,][]{inoue96},
where $m_{\rm e}$ is the electron rest mass, 
$u_{\rm B}=\frac{B^2}{8\pi}$ is the energy density of the magnetic field $B$,
$\sigma_{\rm T}$ is the Thomson cross section,
$R$ is the thickness of the acceleration region 
for electrons with $\gamma_{\rm b}$, 
and $v$ is the downstream flow velocity in the shock frame. 
In this condition, the magnetic field is constrained from $\nu_{\rm c}$
($\propto {\gamma_{\rm b}}^2$) as follows; 
\begin{equation}
B^3 \simeq 
	\frac{27 \pi e m_{\rm e} v^2  c}{\sigma_{\rm T}^2} R^{-2} \nu_{\rm c}^{-1},
\label{eq:param}
\end{equation}
where $e$ is the electron charge.

%====================================================%
\subsection{Comparison with the minimum-energy magnetic field.}
\label{sec:comp-Bme}
\begin{figure}[t]
\plotone{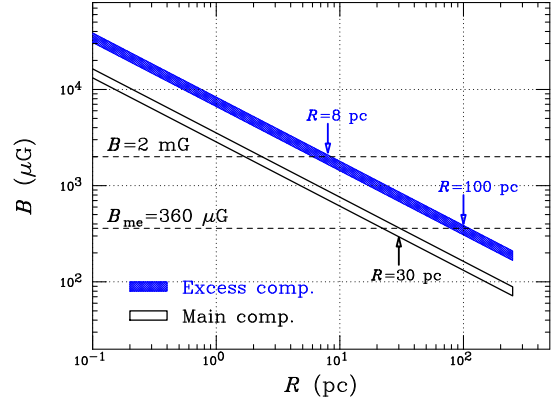}
%{../fig/param/B-R_fit3-1_syserr_0.3c.ps}
\caption{
The relation between the magnetic field $B$ and 
the thickness of the acceleration region $R$ for 
a representative downstream flow velocity of $v = 0.3 c$,  
estimated from $\nu_{\rm c}$ under the cooling break hypothesis.
The acceptable spaces to the main and excess components 
are indicated by the thick solid line and blue hatched area, respectively.
The upper limit on $R$ was estimated from the radio emitting volume 
\citep[$R=250$ pc;][]{wilson01}. 
The horizontal dashed lines indicate the magnetic field strength of
$B_{\rm me} = 360$ $\mu$G and $B=2$ mG.  
The vertical arrows denote the representative region sizes
discussed in \S\ref{sec:comp-Bme}.}
\label{fig:B-R}
\end{figure}

Figure \ref{fig:B-R} shows the acceptable parameter space on the $B$--$R$ plane,
derived from Equation \ref{eq:param}, 
for a representative value of the flow velocity, $v = 0.3 c$ \citep[e.g.,][]{kino04}.
The excess component tends to exhibit a higher magnetic field
due to its lower $\nu_{\rm c}$ value.

In order to infer the physical condition for the main component, 
the minimum energy argument \citep{miley80} is 
adopted as the simplest working hypothesis.
Adopting the radio emitting volume, 
which is equivalent to a sphere with a radius of $250$ pc 
\citep{wilson01,tingay08}, 
the spectral energy distribution shown in Figure \ref{fig:SED} 
yields the minimum-energy magnetic field as $B_{\rm me} = 360$ $\mu$G.
Here, the proton energy density is assumed to be equal to the electron one. 
On the field strength, 
It is suggested that the synchrotron electrons 
corresponding to $\nu_{\rm c}$ for the main component 
are generated within $R \sim 20$--$30$ pc around the shock,
as shown with the upward arrow in Figure \ref{fig:B-R}.
The estimated size is consistent with 
the result from the detailed spectral modeling of this hot spot
without mid-infrared data \citep[$\lesssim 50$ pc;][]{meisenheimer97}. 
 
The first interpretation of the excess component
assumes that the two components originate 
in the different regions within the hot spot,
but with the same magnetic field (i.e., $B_{\rm me} = 360$ $\mu$G).
The electron cooling time scale of the excess component 
($t_{\rm syn} \lesssim 10^3$ yrs) is expected to 
be longer by a factor of $3.6$ than that of the main component,
due to its lower cutoff frequency by a factor of $13$.
This implies that the excess electron population is older than the main one.
Accordingly, 
the excess electrons is expected to be distributed in a wider region 
($R\lesssim 100$ pc, see the right downward arrow in Figure \ref{fig:B-R}) 
than the main ones.

The second and more attractive explanation hypothesizes 
that the excess is produced in a region with a stronger magnetic field. 
A number of theoretical studies 
\citep[e.g.,][]{lucek00,inoue09,fraschetti13,araudo15,ji16} 
proposed that the magnetic field is amplified 
by plasma instabilities, turbulences, or non-linear interactions 
between the field and accelerated particles. 
Based on these studies, 
the amplification by a factor of at least $4$--$6$
is achievable \citep{lucek00}. 
Thus, 
the minimum energy field in the west hot spot of Pictor A
is possible to be magnified up to $B\sim 2$ mG. 
On Figure \ref{fig:B-R},
this field corresponds to the thickness of $R=6$--$8$ pc 
(shown with the left downward arrow). 

With the VLBA image, \citet{tingay08} resolved the hot spot of Pictor A 
into sub structures with a size of $\gtrsim 10$ pc.
Summed over these sub structures, 
the radio flux density was measured as $F_{\nu} = 121 \pm 6$ mJy 
at $1.67$ GHz.  
They interpreted that 
these sub structures correspond to the recent shocks 
with a higher energy density (i.e, a stronger magnetic field)
than the mean energy density over the hot spot. 
Interestingly, 
the size of the excess component is comparable to 
%(or slightly smaller than) 
those of these sub structures. 
In addition, 
the $1.67$ GHz flux density of the excess component,
which is estimates from the double cutoff-PL fitting
as $F_{\nu} \sim 50$--$500$ mJy, 
agrees with that of the sub structures. 
Therefore, it is natural to arrive at the scenario that 
the acceleration site of the excess component is 
some of these sub structures with a strengthened magnetic field.

%----------% x
% figure 4 % x
%----------% x

%====================================================%
\subsection{Relation to the origin of the X-ray spectrum}
\label{sec:X-ray}
The X-ray spectrum measured with \chandra\ 
\citep[the photon index of $\Gamma = 2.07 \pm 0.11$;][]{wilson01}
is reported to be significantly steeper than the synchrotron radio one. 
It is widely known that this inconsistency 
between the radio and X-ray spectral slope makes unfeasible 
the simple one-zone synchrotron+SSC interpretation 
for the wide band spectral energy distribution from the object. 
\citet{tingay08} proposed that 
fresh electrons accelerated in the recent shocks,
which are traced by the VLBA sub structures,
produce the observed X-ray spectrum via synchrotron radiation.
Their hybrid synchrotron model,
where the radio and X-ray photons originate in the whole hot spot 
and compact regions respectively, successfully reproduced 
the overall spectral energy distribution. 
From the model, the synchrotron peak of the compact component 
is expected to be located in the X-ray band. 
The X-ray spectral break suggested in the deep \chandra\ observation 
\citep{hardcastle16} possibly supports this explanation. 

By modifying the scenario by \citet{tingay08},
the mid-infrared excess revealed with \wise\ and \spitzer\ is suggested 
to request three different populations of synchrotron electrons 
in the west hot spot of Pictor A; 
two of which are localized in the compact sub structures. 
Assuming the magnetic field in these compact regions is magnified up to 
$B = 2$ mG, as discussed in \S\ref{sec:comp-Bme}, 
the synchrotron cooling time scale of the X-ray electrons is 
evaluated as $t_{\rm syn} \sim 1$ yrs $E_{\rm keV}^{-0.5}$ 
where $E_{\rm keV}$ is the synchrotron photon energy in the unit of keV.
This is significantly shorter than that of the mid-infrared excess, 
$t_{\rm syn} \sim 75$ yrs $(\nu_{c}/5.5\times10^{13} {\rm~Hz})^{-0.5}$.
Therefore, one of the plausible ideas is that 
among the VLBA sub structures, 
the younger ones are the origin of the X-ray spectrum,
while the mid-infrared excess are radiated from the older ones. 

It is claimed that 
the compact structures with an amplified magnetic field are a transient feature, 
since they rapidly expand to achieve pressure balance with the ambient region
\citep{tingay08}. 
As a result, 
the emission from these compact regions are expected to be variable 
with a short time scale. 
This is possibly confirmed by a series of \chandra\ observations 
of the west hot spot of Pictor A,
in which the X-ray flux decayed by $\sim 10$\% 
in a time scale of $3$ months \citep{hardcastle16}.
The variability implies that 
the core of the X-ray emitting region is possible to be in a sub-pc scale. 
Between the \wise\ and \spitzer\ observations separated by $\gtrsim 5$ yrs, 
no significant change was indicated in the mid-infrared flux, 
when all the photometric uncertainties are taken into consideration,
including the \wise\ systematic errors 
($\sigma_{\rm sys}$ in Table \ref{tab:catalog}) 
and the \spitzer\ one \citep[typically $3$\%;][]{carey12}. 

%====================================================%
\subsection{Prospects}
\label{sec:prospects}
The striking implication from the close investigation into 
the synchrotron spectrum of the west hot spot of Pictor A is 
that the radio and mid-infrared spectra are of different origin
from each other, as is visualized in the bottom panel in Figure \ref{fig:SED}.
If this is the case in other objects, 
the mid-infrared emission is inferred to dominate over the radio one 
in some particular condition,
although what controls the relative importance 
between the radio and mid-infrared emission remains unsettled.
Currently, hot spots,
that are bright in the mid-infrared band and instead relatively weak 
in the radio band, have not yet known.
The all-sky survey data with \wise\ has a great potential 
to search for such ``mid-infrared hot spots."

The multi-zone interpretation presented above
is expected to be reinforced by precisely measuring the synchrotron spectrum 
in between the radio and mid-infrared frequencies. 
However,
the far-infrared flux of the west hot spot of Pictor A
estimated from the \wise\ result, $\sim35$ mJy at $90$ $\mu$m, 
is below the detection limit of previous far-infrared all-sky surveys,
including the \akari\ all-sky survey catalog
\citep[$0.43$ Jy at $90$ $\mu$m;][]{yamamura12}.
Future observations with the next-generation infrared observatory {\itshape SPICA} 
\citep[Space Infrared Telescope for Cosmology and Astrophysics;][]{roelfsema17} 
and ALMA are of prime importance. 
In addition, 
multi-epoch observations with the James Webb Space Telescope
in the near future 
are useful to search for mid-infrared flux variation. 
This will constrain the regions size
and thus reconfirm the sub-structure explanation
for the mid-infrared excess from the west hot spot of Pictor A.

\acknowledgments
The authors are grateful to the anonymous referee 
for her/his supportive and constructive comments.
This study makes use of data products from the \wise\ and NEOWISE projects.
The unpublished ATCA data of Pictor A is kindly provided by Dr. E. Lenc. 
This research is supported by the MEXT/JSPS KAKENHI 
Grants (JP24103002, JP25109001, and JP26247030).

%----------------%
% Bibliography %
%----------------%

\end{document}